\documentclass[doublecol]{epl2} 

\newcommand{\be}{\begin{equation}}
\newcommand{\ee}{\end{equation}}
\newcommand{\bea}{\begin{eqnarray}}
\newcommand{\eea}{\end{eqnarray}}

\title{Heterogeneous pair approximation for voter models on networks}
\shorttitle{Heterogeneous pair approximation for voter models on networks} 

\author{Emanuele Pugliese\inst{1,2} \and Claudio Castellano\inst{3}}
\shortauthor{E. Pugliese \etal}

\institute{                    
  \inst{1} Dipartimento di Fisica, ``Sapienza'' Universit\`a di Roma,
P.le Aldo Moro 2, I-00185 Roma, Italy \\
  \inst{2} S. Anna School of Advanced Studies, P.za Martiri della
Libert\`a 33, I-56100 Pisa, Italy \\
  \inst{3} SMC, INFM-CNR and
Dipartimento di Fisica, ``Sapienza'' Universit\`a di Roma,
P.le Aldo Moro 2, I-00185 Roma, Italy
}
\pacs{87.23.Ge} {Dynamics of social systems}
\pacs{64.60.aq} {Networks}
\pacs{05.40.-a} {Fluctuation phenomena, random processes, noise, and Brownian motion}

\abstract{
For models whose evolution takes place on a network it is often necessary
to augment the mean-field approach by considering explicitly the
degree dependence of average quantities (heterogeneous mean-field).
Here we introduce the degree dependence in the pair approximation
(heterogeneous pair approximation) for analyzing voter models on
uncorrelated networks.
This approach gives an essentially exact description of the dynamics,
correcting some inaccurate results of previous approaches. 
The heterogeneous pair approximation introduced here can be applied
in full generality to many other processes on complex networks.
}

\begin{document}

\maketitle

\section{Introduction}

The voter model is one of the simplest possible models of nonequilibrium
dynamics, as witnessed by the number of different contexts where it has been 
considered, ranging from probability to theoretical ecology
and heterogeneous catalysis~\cite{Liggett85,Durrett96,Castellano09}.

Each node of a graph is endowed with a binary variable (a spin) $s=\pm 1$.
At each time step a node and one of its neighbors are chosen at random
and the first node becomes equal to the second. This microscopic dynamics
gives rise to a nontrivial macroscopic
phenomenology~\cite{Castellano09} with the formation of correlated domains
that tend to grow in time.
In finite systems this eventually leads to one of the two possible ordered
states (consensus) with either all spins up ($+1$) or down ($-1$).
In the thermodynamic limit full order is reached only in euclidean
dimension $d \le 2$.

From the point of view of statistical physics voter dynamics has the very
interesting feature that its overly simple rules allow
for analytical treatment in many cases.
On euclidean lattices many exact results
are available~\cite{Krapivsky92, Frachebourg96}.
In the past few years, in the framework of the huge interest around
complex networks and dynamics on top of
them~\cite{Dorogovtsev02, Newman03, Barrat08}, the effect of nontrivial
topologies on voter dynamics has been explored~\cite{Castellano03,Vilone04,Sood05,Suchecki05,Castellano05,Castellano05b,Suchecki05b, Sood08,Blythe08,Baronchelli
08}.
One interesting effect of the
presence of a disordered connectivity pattern is that slightly different
definitions of the model, perfectly equivalent on lattices,
become inequivalent on networks.
In the {\em direct voter} case, one picks up a vertex at random and this
imitates a randomly chosen neighbor.
In the {\em reverse voter} dynamics~\cite{Castellano05} or
{\em invasion process} the order of selection is opposite:
it is the neighbor that imitates the node selected first.
An intermediate case is the {\em link-update dynamics}~\cite{Suchecki05},
where one has to choose at random first a link and then which of the two
nodes at the end of the link imitates the other.
Only the link-update rule preserves a crucial property of voter dynamics
on lattices, conservation on average of magnetization~\cite{Suchecki05},
whereas other quantities are conserved in the two other 
cases~\cite{Sood05, Castellano05}.
On regular geometries the three types of dynamics coincide, but when
the topology is strongly heterogeneous important differences arise:
on scale-free networks~\cite{Barabasi99}
the scaling of average consensus time $T$
with the system size $N$ changes depending on which of the three
versions is considered.

Theoretical progress on the understanding of these issues has been made
possible by the application of a heterogeneous mean-field
approach~\cite{Sood05, Castellano05, Sood08}.
The basic idea is to define the densities $\sigma_k$ of up spins restricted to 
nodes of degree $k$ and to write down evolution equations for them.
In this way it is possible to predict that for the direct voter
the average time $T$ to reach consensus grows sublinearly
as a function of the system size $N$
if the exponent $\gamma$ of the tail of the degree distribution is smaller
than $3$, while it is linear for $\gamma>3$. Instead, for the reverse and
link-update versions of the dynamics the growth is linear for
any $\gamma>2$.
Very recently, the theoretical picture for the voter model
has been further refined by the application of an improved
mean-field approach, a homogeneous pair approximation~\cite{Vazquez08},
taking into account dynamical correlations among nearest neighbors.
This approach allows to follow in detail the temporal evolution
of the system and to compute with remarkable accuracy even the prefactor
of the law that expresses the average consensus time as a power of $N$.

Despite these achievements, some open problems remain, calling
for further improvement.
The first has to do with the prediction of a transition
for average degree $\mu$ equal to 2.
According to Ref.~\cite{Vazquez08}, while for $\mu>2$
consensus is attained via a quasi-stationary intermediate state over
a long temporal scale, when $\mu<2$ order is reached exponentially fast.
Contrary to this prediction, numerical simulations on an Erd\"os-Renyi
graph do not display any singular behavior in correspondence of
$\mu=2$~\cite{Vazquez08}.
Another issue that requires a clarification has to do with the use
of the same approach for reverse voter dynamics. As it will be
explained below, the application is straightforward, but the
agreement with numerical simulations is not complete.

In this paper we go beyond the homogeneous pair approximation
introduced in Ref.~\cite{Vazquez08} by 
allowing correlations between nearest neighbor nodes to depend
on their degree (heterogeneous pair approximation).
In practice we lift the assumption of Ref.~\cite{Vazquez08} that the
probability for a link between two nodes to be active 
(i.e. to connect nodes in different spin state) does not depend
on the degree of the nodes.
This is analogous to the introduction of the heterogeneous mean-field
approach~\cite{Pastor-Satorras01, Barrat08} necessary for dealing with the
SIS model (and many others) on scale-free networks.
With the heterogeneous pair approximation we find, for the direct
voter dynamics, results in general very close to those obtained with
the homogeneous pair approximation but we get rid of the spurious
transition at $\mu=2$, in agreement with computer simulations.
Also for the reverse voter and link-update dynamics we obtain results
in very good agreement with numerics, showing that the heterogeneous
pair approximation fully captures the behavior of voter models.
These successful results call for the application of the heterogeneous pair
approach to other nontrivial dynamical processes on networks, 
including also epidemiological models or evolutionary
games~\cite{Lieberman05}.
While the detailed implementation will differ depending on the case,
the idea is completely general and may lead to substantial progress
with respect to the heterogeneous mean-field approach.

\section{Problems with the homogeneous pair approximation}
\label{need}

\begin{figure}
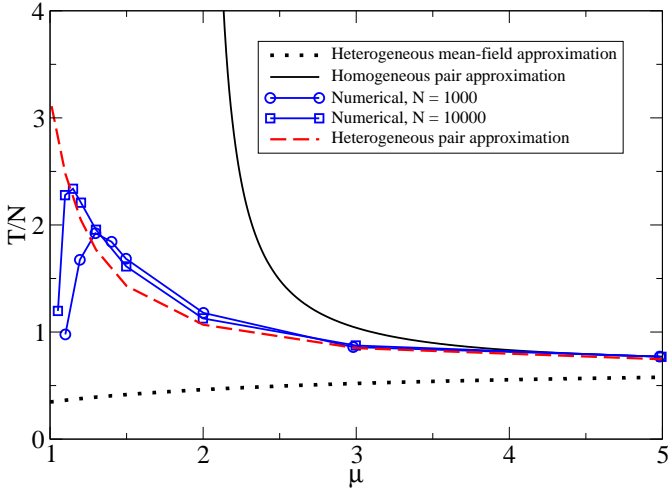

\begin{center}
\onefigure[width=\columnwidth]{T_ER.eps}
\end{center}
\caption{Average time to reach full consensus for direct voter dynamics
on an Erd\"os-Renyi graph, as a function of the average degree $\mu$,
starting from a symmetric uncorrelated initial configuration 
$\sigma_k(t=0)=1/2$.
The curve for the heterogeneous mean-field approach is obtained using
the formula $T/N = \mu^2/\mu_2 \ln(2)$ from Ref.~\cite{Sood05}.
The curve for the homogeneous pair approximation is obtained using
Eq.~(\ref{T_voter}). The curve for the heterogeneous
pair approximation is obtained using Eq.~(\ref{Tdirect}).
For finite $N$ and $\mu$ very close to 1, the network is
separated in many nonextensive disconnected components. This is the
origin of the discrepancy between numerical results and the
theoretical prediction in an interval that shrinks to zero as the
system size diverges.
Each point is the average over 1000 realizations.}
\label{T_ER}
\end{figure}

The homogeneous pair approximation introduced in Ref.~\cite{Vazquez08} is
based on the derivation of the equation of motion for the quantity $\rho$,
the fraction of active links, i. e. edges
connecting nodes in opposite spin state.
In the derivation, it is assumed that the probability
for a link in the system to be active does not depend on the
degree of the nodes it connects, and is equal to $\rho$.
Under this hypothesis the equation of motion for $\rho$ is derived,
involving the magnetization $m$ and the first moment of the degree
distribution $\mu$. 
As mentioned before, 
the behavior predicted changes dramatically as a function of $\mu$.
For $\mu<2$, $\rho$ goes to zero exponentially fast in time, indicating
that a fully ordered configuration (consensus) is quickly reached: the
average time $T$ to reach consensus does not depend on the
system size $N$ so that $T/N \to 0$.
For $\mu>2$ instead it is predicted that, over a very short temporal
scale, a quasi-stationary state is reached,
characterized by a density of active links
\be
\rho^S = \frac{\mu-2}{2(\mu-1)} (1-m^2).
\label{rho_s_voter}
\ee
The superscript $S$ indicates, throughout the Letter,
quantities in the quasi-stationary state.
The eventual behavior of the system consists in the erratic
variation of $m$ and $\rho^S$ that jointly fluctuate in time while
obeying Eq.~(\ref{rho_s_voter}), until a fluctuation leads to the absorbing
state $|m|=1$, $\rho=0$, corresponding to full consensus.
Other dynamical observables, describing the ordering process, are
expressed using Eq.~(\ref{rho_s_voter}), as for example the average
consensus time that, for uncorrelated initial conditions
with magnetization $m=0$, is
\be
T =  \frac {\mu^2 N}{2 \rho^S(m=0) \mu_2} \ln(2) = 
     \frac {(\mu-1) \mu^2 N}{(\mu-2) \mu_2} \ln(2) ,
\label{T_voter}
\ee
where $\mu_2$ is the second moment of the degree distribution.
While Eq.~(\ref{rho_s_voter}) agrees with
remarkable precision with the outcome of numerical simulations
for large values of $\mu$, 
significant discrepancies arise on many types of networks
when $\mu$ gets close to $2$.
For example, considering as a substrate an Erd\"os-Renyi graph,
the divergence of $T/N$ for $\mu \to 2$ predicted by
Eq.~(\ref{T_voter}) is not
seen in numerical simulations~\cite{Vazquez08} (Fig.~\ref{T_ER}).
This indicates that, despite the very good performance for large values
of $\mu$, the homogeneous pair approximation of Ref.~\cite{Vazquez08}
qualitatively fails in the limit $\mu \to 2$.

\begin{figure}
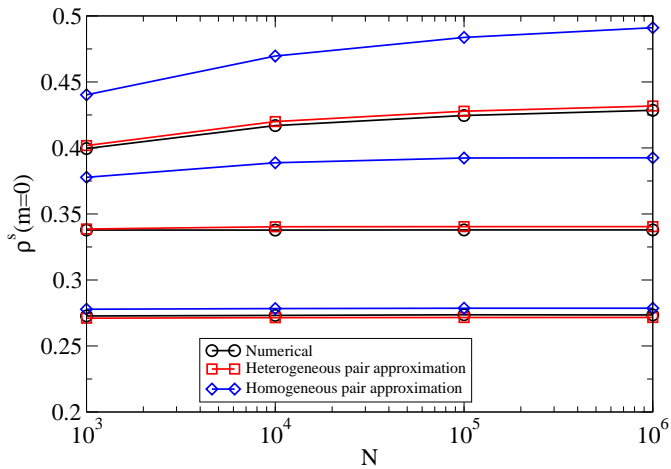

\begin{center}
\onefigure[width=\columnwidth]{comparison_rvoter2.eps}
\end{center}
\caption{Reverse voter dynamics: 
quasi-stationary value $\rho^S(m=0)$ as a function of the system
size for three different values of $\gamma$ (top to bottom, $\gamma=2.5$,
$\gamma=4$, $\gamma=8$).}
\label{comparison_rvoter2}
\end{figure}

The same indication emerges when the approach is applied to the reverse
voter dynamics.
For the quasi-stationary value of the density of active links it is
straightforward to obtain
\be
\rho^S = \frac{\mu_2 - 2 \mu}{2(\mu_2-\mu)} (1-m^2).
\label{rho_s_rvoter}
\ee
This formula coincides, as expected, with the result for the direct voter
model [Eq.~(\ref{rho_s_voter})] when the degree distribution is a
delta-function so that $\mu_2=\mu^2$.
However, a careful comparison with numerics shows that
Eq.~(\ref{rho_s_rvoter}) is not correct when $\mu_2 \gg \mu^2$
(see Figure~\ref{comparison_rvoter2}).
The parabolic dependence on magnetization remains (not shown), but the
value for $m=0$ predicted by Eq.~(\ref{rho_s_rvoter}) (diamond symbols) is
considerably larger than what simulations give (circles),
unless $\gamma$ is very large.
Here, as in the rest of the paper unless specified otherwise, 
we compare the results of the theoretical treatment with simulations
performed on the uncorrelated configuration model (UCM)~\cite{Catanzaro05},
that generates graphs with power-law degree distribution
$P_k \sim k^{-\gamma}$ and upper bound of the degree distribution
equal to $N^{1/2}$, so that automatically there are 
no self-loops nor double-links . The lower-cutoff of the degree distribution
is $3$.

\begin{figure}
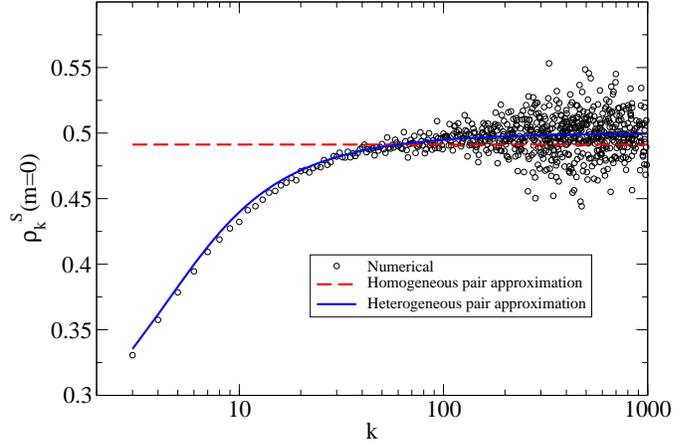

\begin{center}
\onefigure[width=\columnwidth]{rho_k_vs_k_rvoter.eps}
\end{center}
\caption{Plot of the probability that a link connected to a node of degree
$k$ is active in the quasi-stationary state (and for $m=0$) for reverse voter
dynamics taking place on an uncorrelated
network with $\gamma=2.5$ and size $N=10^6$. The black symbols are the
results of a numerical simulation. The dashed red line is the prediction
of the homogeneous pair approximation, Eq.~(\ref{rho_s_rvoter}), the
blue solid line is the prediction of the heterogeneous pair approximation.}
\label{rho_k_vs_k_rvoter}
\end{figure}

\begin{figure}
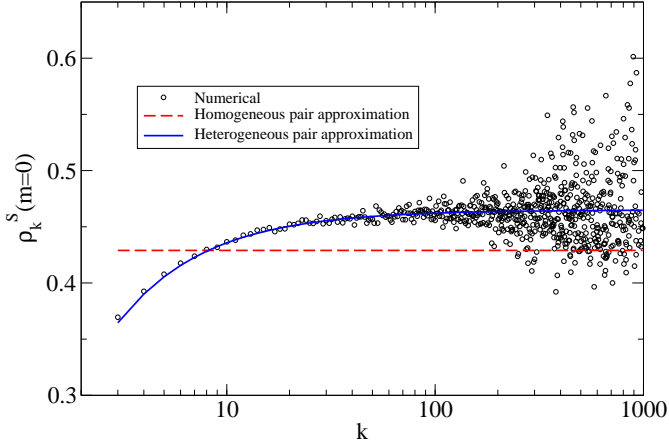

\begin{center}
\onefigure[width=\columnwidth]{rho_k_vs_k_voter.eps}
\end{center}
\caption{Plot of the probability that a link connected to a node of degree
$k$ is active in the quasi-stationary state (and for $m=0$) for direct voter
dynamics taking place on an uncorrelated
network with $\gamma=2.5$ and size $N=10^6$. The black symbols are the
results of a numerical simulation. The dashed red line is the prediction
of the homogeneous pair approximation, Eq.~(\ref{rho_s_voter}), the
blue solid line is the prediction of the heterogeneous pair approximation.}
\label{rho_k_vs_k_voter}
\end{figure}

The origin of the discrepancies in Figs.~(\ref{T_ER})
and~(\ref{comparison_rvoter2}) is not hard to identify.
It is sufficient to evaluate numerically the quantity $\rho^S_k$,
defined as the fraction of neighbors of a node of degree $k$
that are in a different spin state in the quasi-stationary state,
and plot it as a function of $k$ (Fig.~\ref{rho_k_vs_k_rvoter}).
While the approach of Ref.~\cite{Vazquez08} assumes this quantity
to be the same for all nodes, a substantial dependence on $k$ shows up.
Interestingly, a numerical check indicates that also for direct voter
dynamics $\rho^S_k$ has a sizeable dependence on the degree
(Fig.~\ref{rho_k_vs_k_voter}).
Based on this evidence, we now generalize the pair approximation
approach by considering $\rho_{k,k'}$, i.e. allowing the density
of active links to depend explicitly on the degree on the connected nodes.
From $\rho_{k,k'}$ it is possible to determine all other quantities
of interest.
The density $\rho_k$ of active links connected to a node of degree $k$
is $\rho_k=\sum_{k'} Q_{k'} \rho_{k,k'}$, where $Q_k = P_k k/\mu$
is the degree distribution of the neighbours of a randomly chosen node.
The total density of active links $\rho$ is likewise given by
$\rho=\sum_k Q_k \rho_k = \sum_{k,k'} Q_k Q_{k'} \rho_{k,k'}$.

\section{Heterogeneous pair approximation for the direct voter model}
\label{voter}

The first goal is to determine the equation of motion for $\rho_{k,k'}$.
This quantity is modified if the flipping node has degree
$k$ and one of its neighbours has degree $k'$ (or vice versa).
Let us assume that the flipping (first selected) node has degree $k$ and call 
$k''$ the degree of the copied (second selected) node.
It is useful to consider separately the two cases where $k'' \ne k'$ or
$k'' = k'$.

\begin{figure}
\begin{center}
\onefigure[width=\columnwidth]{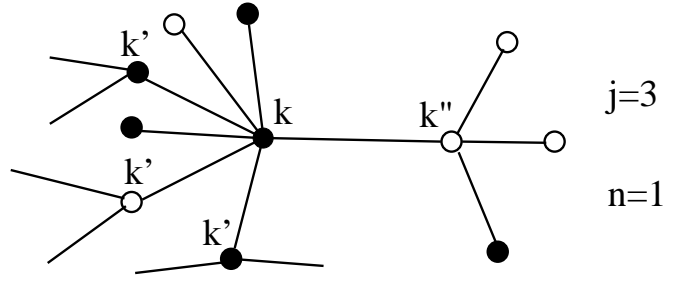}
\end{center}
\caption{Direct voter dynamics:
Illustrative representation of a
configuration contributing to $d\rho_{3,7}/dt$.
The first node selected has degree $k=7$ and the second has degree
$k''=4$, different from $k'=3$. Filled (empty)
circles denote spins in state $s=+1$ ($s=-1$).
The variation to the density $\rho_{k,k'}$ is due to the $j=3$
links that connect the flipping node to nodes of degree $k'$.
Before the flipping of the first node selected, $n=1$ of the links
are active and $j-n=2$ are inactive.
After the flipping the state of links is reversed and the
variation in the number of active links between nodes of degree
$k$ and $k'$ is $j-2n=1$.}
\label{hpafigure}
\end{figure}

In the first case the variation $\Delta \rho_{k,k'}$ for a single
dynamical step (occurring over a time $\Delta t = 1/N$) is determined as
follows. The probability that a node in state $s$ and degree $k$
flips is given by
the probability $P_k$ that the first node selected has degree $k$ times
the probability $\sigma(s)$ that it is in state $s$,
times the probability $Q_{k''}$ that the second
has degree $k''$ multiplied by the probability $\rho_{k,k''}/[2 \sigma(s)]$
that the link connecting the two is active. 
One has then to multiply this quantity by the associated variation of the
fraction of active links between $k$ and $k'$ (Fig.~\ref{hpafigure}).
Among the $k-1$ other links
of the flipping node, the number of those connecting to a node of degree $k'$
will be $j$ distributed according to a binomial $R(j,k-1)$ with probability
of the single event equal to $Q_{k'}$. In their turn, only $n$ out of these
$j$ links will be active, with $n$ binomially distributed with single event
probability $\rho_{k,k'}/[2 \sigma(s)]$. Finally one has to multiply by the
variation of $\rho_{k,k'}$ when $n$ out of $j$ links go from active to
inactive as a consequence of the flipping of the node in $k$.
This is given by the variation of the number of active links $[(j-n)-n]$
divided the total number of links between
nodes of degree $k$ and $k'$ ($N \mu Q_k Q_k'$).
One has then to sum over $k'' \ne k'$, $s$, $j$ and $n$, obtaining
\bea
\label{Deltarho1}
\Delta \rho_{k,k'} & = & P_{k} \sum_s \sigma(s) 
\sum_{k''\ne k'} Q_{k''} \frac{\rho_{k,k''}}{2\sigma(s)} \cdot \\ \nonumber
& & \sum_{j=0}^{k-1} R(j,k-1) \sum_{n=0}^j B(n,j) \frac{j-2n}{N \mu Q_k Q_{k'}}.
\eea
By performing explicitly the summations (and using 
$\sum_s 1/\sigma(s)=4/(1-m^2)$) the formula becomes
\be
\frac{\Delta \rho_{k,k'}}{\Delta t} = \sum_{k''\ne k'} Q_{k''} \rho_{k,k''}
\frac{(k-1)}{k}\left(1-\frac{2}{1-m^2} \rho_{k,k'} \right).
\ee
When  $k''=k'$, the value of $\Delta \rho_{k,k'}$ is similar to
Eq.~(\ref{Deltarho1}) with (obviously) $Q_{k'}$ instead of $Q_{k''}$,
no sum over $k''$, and in the numerator of the last factor
$j+1-(n+1)-(n+1)=j-2n-1$, because
there are $j+1$ links to nodes of degree $k$, $n+1$ of which are active in
the initial state and inactive in the final. Summing up the two contributions
and adding the symmetric terms with $k$ and $k'$ swapped, we get
\bea
\label{drhokk'_voter}
\frac{d\rho_{k,k'}}{dt}=\rho_k\frac{k-1}{k}+\rho_{k'}\frac{k'-1}{k'}+\\ \nonumber
-\rho_{k,k'}\left(\frac{1}{k}+\frac{1}{k'}+\frac{2\rho_k}{1-m^2}\frac{k-1}{k}+\frac{2\rho_{k'}}{1-m^2}\frac{k'-1}{k'}\right).
\eea
From this equation, by summing over $k$ and $k'$ one obtains
the equation of motion for the total density $\rho$
\bea
\label{drho_voter}
\frac{d\rho}{dt} & = &\sum_{k,k'} Q_k Q_{k'}\frac{d\rho_{k,k'}}{dt} \nonumber \\
&=&2\left(\rho-2\frac{\pi}{\mu}\right)-\frac{4}{1-m^2}\left(\rho_2^*-\frac{\pi_2^*}{\mu}\right),
\eea
where $\pi = \sum_k P_k \rho_k$, $\pi_2^* = \sum_k P_k \rho_k^2$
and $\rho_2^* = \sum_k Q_k \rho_k^2$.
It is easy to see that, if one assumes $\rho_k = \rho$, i.e. that
the density does not depend on the degree (and as a consequence
$\pi=\rho$, $\rho_2^* = \pi_2^* = \rho^2$), Eq.~(\ref{drho_voter})
coincides with the analogous equation of Ref.~\cite{Vazquez08}.

After a time scale of order unity the quasi-stationary state is established.
From $d\rho_{k,k'}/dt=0$ we obtain
\be
\label{rhokk'_s_voter}
\rho^S_{k,k'}=\frac{\rho^S_k\frac{k-1}{k}+\rho^S_{k'}\frac{k'-1}{k'}}{\frac{2}{1-m^2}\left(\rho^S_k\frac{k-1}{k}+\rho^S_{k'}\frac{k'-1}{k'}\right)+\frac{1}{k}+\frac{1}{k'}}
\ee
From this equation it is evident that all $\rho^S_{k,k'}$ are proportional to
$1-m^2$.
Imposing the consistency condition $\rho_k=\sum_{k'}Q_{k'}\rho_{k,k'}$,
we have a set of coupled equations for $\rho^S_k$ for all $k$.
The numerical iterative solution of this set of equations allows
the determination of the quasi-stationary values of all $\rho^S_k$ and hence
of all $\rho^S_{k,k'}$.

Fig.~\ref{rho_k_vs_k_voter} shows that the heterogeneous
pair approximation captures extremely well the detailed
dependence of $\rho^S_k$ on $k$.
Also the comparison of the quasi-stationary value of $\rho^S$ (recovered
by summing  Eq.~(\ref{rhokk'_s_voter}) over $k$ and $k'$)
with simulations for several values
of $\gamma$ and $N$ indicates an excellent agreement (not shown).

The knowledge of the $\rho^S_k$ allows to determine also the consensus
time $T$ to reach a fully ordered configuration. Following the
formalism of Ref.~\cite{Sood08}, using the backward Kolmogorov equation,
one simply needs to express the transition probabilities $R_k$ and $L_k$
in terms of the $\rho_k$.
In the quasi-stationary state the probability $R^S_k$ that the number
of nodes of degree $k$ in state $s=+1$ increases by $1$ is
\be
R^S_k = P_k (1-\sigma_k) \sum_{k''} Q_{k''} \rho^S_{k,k''}/[2 (1-\sigma_k)] =
P_k \frac{\rho^S_k}{2}.
\ee
The probability $L_k$ that the number is reduced is the same, $L^S_k=R^S_k$.
Using these expressions one obtains
\be
T = - \tau(N)
\left[(1-\omega) \ln(1-\omega) -\omega \ln \omega \right]
\label{Tdirect}
\ee
where $\omega = \sum_k Q_k \sigma_k(t=0)$ is conserved by the dynamics
and
\be
\tau(N)=\frac{N\mu^2}{2\sum_k P_k k^2 \rho^S_k(m=0)}
\label{taudirect}
\ee
sets the temporal scale over which consensus is reached.
Notice that with the mean-field assumption, $\rho^S_k(m=0)=1/2$,
Eq.~(\ref{taudirect}) coincides with the result of Sood and Redner
$\tau(N)=N \mu^2/\mu_2$. With the homogeneous pair approximation
$\rho^S_k(m=0)=(\mu-2)/[2(\mu-1)]$ it returns the result of Vazquez
and Egu\'iluz~\cite{Vazquez08}.
With Eq.~(\ref{Tdirect}), valid for any degree distribution,
we have computed the value of the ratio $T/N$
for an Erd\"os-Renyi graph of average degree $\mu$. The plot in
Fig.~\ref{T_ER} presents the comparison of this value with the prediction
of previous approaches and with the outcome of numerical simulations.
The singular behavior for $\mu=2$ is an artifact of the homogeneous
pair approximation, that is removed by the heterogeneous pair approximation,
leading instead to a smooth behavior around $\mu=2$
and a very good agreement between theory and simulations.

\section{Heterogeneous pair approximation for the reverse and link-update voter dynamics}
\label{reverse}

The application of the heterogeneous pair approximation to the reverse
voter model proceeds along the same lines of the direct dynamics.
The only difference is that the factor $P_{k} Q_{k''}$ is replaced by
$P_{k''} Q_{k}$, because the role of the flipping and the copied nodes
is swapped. As a consequence, the formula for the quasi-stationary value
of $\rho_{k,k'}$ is slightly modified
\bea
\label{rhokk'_s_rvoter}
\rho^S_{k,k'}=\frac{\pi^S_k(k-1)+\pi^S_{k'}(k'-1)}
{\frac{2}{1-m^2}\left[\pi^S_k(k-1)+\pi^S_{k'}(k'-1)\right]+\mu/k+\mu/k'}.
\eea
where $\pi_k = \sum_{k'} P_{k'} \rho_{k,k'}$.
Much in the same way as for the direct voter dynamics a set of consistency
equations for $\pi^S_k$ is solved iteratively. The values of $\pi^S_k$,
inserted into Eq.~(\ref{rhokk'_s_rvoter}) provide $\rho^S_{k,k'}$,
from which in turn $\rho^S_k$ and $\rho^S$ can be computed.
Fig.~\ref{rho_k_vs_k_rvoter} shows that the agreement with
the values of the $\rho^S_k$ obtained numerically is very good.
Moreover, Fig.~\ref{comparison_rvoter2} shows that the quasi-stationary value
of $\rho^S$ is in fully satisfactory agreement with numerics for several
values of $\gamma$ and $N$.

\begin{figure}
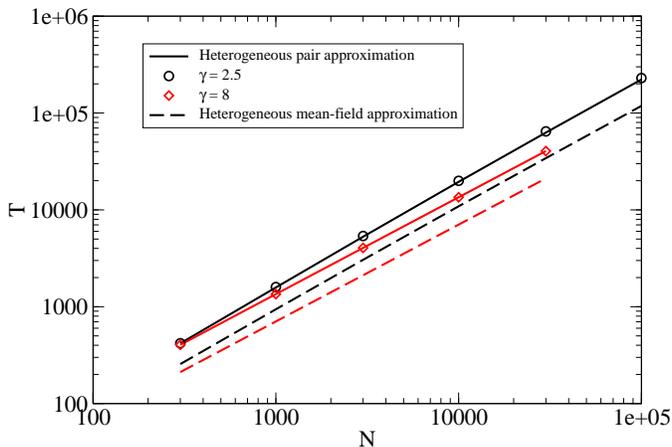

\begin{center}
\onefigure[width=\columnwidth]{Tcomparison.eps}
\end{center}
\caption{Reverse voter dynamics: Plot of the consensus time $T$ as a 
function of the system size $N$ for two values of gamma. Filled symbols
are for numerical results, solid lines are the theoretical predictions
of Eq.~(\ref{T}), dashed lines are the mean-field
predictions of Ref.~\cite{Castellano05}.}
\label{Tcomparison}
\end{figure}

Also in this case the consensus time is computed using the
backward Kolmogorov equation for $T$~\cite{Sood08}. The
transition probabilities in the quasi-stationary state are obtained
from those of the direct case with the replacement
$P_{k} Q_{k''} \to P_{k''} Q_{k}$:
\bea
R^S_k & = & Q_k \frac{\pi^S_k}{2} \\ \nonumber
L^S_k & = & R^S_k.
\eea
Using these quantities and the fact~\cite{Castellano05, Sood08} that
the dynamics conserves the quantity
$\omega_{-1} = \sum_k P_k \sigma_k k^{-1}/\mu_{-1}$, 
where $\mu_{-1} = \sum_k P_k/k$, one obtains
\be
T = -\tau(N)
 \left[\omega_{-1} \ln(\omega_{-1})
+ (1 -\omega_{-1}) \ln(1-\omega_{-1}) \right],
\ee
and the temporal scale is
\be
\tau(N) = \frac{N \mu \mu^2_{-1}} {2 \sum_k \frac{P_k}{k} \pi^S_k(m=0)}.
\ee
For uncorrelated initial conditions $\sigma_k(t=0)=1/2 \to \omega_{-1} =1/2$,
\be
T = \tau(N) \ln(2).
\label{T}
\ee
The comparison of the predictions of Eq.~(\ref{T}) with numerical results
is displayed in Fig.~\ref{Tcomparison} and again confirms the correctness
of the theoretical treatment.

For link-update dynamics, all calculations proceed exactly as in the direct
voter case, now with the replacement of $P_k Q_{k''}$ with $Q_k Q_{k''}$.
It is straightforward to obtain
\be
\label{rhokk'_s_lu}
\rho^S_{k,k'}=\frac{\rho^S_k(k-1)+\rho^S_{k'}(k'-1)}
{\frac{2}{1-m^2}\left[\rho^S_k(k-1)+\rho^S_{k'}(k'-1)\right]+2}.
\ee
The consistency condition $\rho_k=\sum_{k'} Q_{k'} \rho_{k,k'}$ is solved
iteratively.
Once again, the values of $\rho^S_k$ and $\rho^S_{k,k'}$ obtained in this way
are in very good agreement with numerics (not shown).
The transition probabilities for the Kolmogorov backward equation are
$R^S_k = L^S_k = Q_k \rho^S_k/2$, that yield, for the consensus time
\be
T = \tau(N)  \left[\omega_0 \ln(\omega_0)
+ (1 -\omega_0) \ln(1-\omega_0) \right],
\ee
where the conserved quantity $\omega_0$ is $\sigma(s=+1)$,
the total density of nodes in state $s=+1$ and $\tau(N)=N/[2 \rho^S(m=0)]$.
Notice that using $\rho^S(m=0)=1/2$ one recovers
the mean-field result $\tau(N)=N$,
i.e. the independence of consensus time of network features~\cite{Castellano05}.
However, since $\rho^S(m=0)$ actually depends on the network,
this independence is just an artifact of the mean-field approach.

\section{Conclusions}
\label{conclusions}

In this Letter, we have introduced a heterogeneous pair approximation
to deal with the ordering dynamics of voter models on networks.
This approach allows to determine, with great precision,
not only the scaling but also the prefactors of the relevant dynamical
observables (quasi-stationary density of active links, consensus time).
The heterogeneous pair approximation thus constitutes a very good
approximate theory for voter models on uncorrelated networks.
The only point that remains to be understood is the origin of the very small
discrepancy (at most of the order of 1\%) that still exists between numerical 
and analytical results.

While the treatment presented here is valid for uncorrelated networks, generalization
to correlated nets is straightforward.
In particular, if $P(k'|k)$ is the probability that a neighbor of 
a randomly chosen node of degree $k$ has degree $k'$, the stationarity
conditions for the three types of dynamics, 
Eqs.~(\ref{rhokk'_s_voter}),~(\ref{rhokk'_s_rvoter}) and~(\ref{rhokk'_s_lu}),
remain the same, provided 
$P(k'|k)$ replaces its expression $Q_k$ valid in the uncorrelated case,
in the definitions of $\rho_k$ and $\pi_k$:
$\rho_k=\sum_{k'} P(k'|k) \rho_{k,k'}$,
$\pi_k=\sum_{k'} \mu P(k'|k)/k' \rho_{k,k'}$. While the generalization of the
analytical approach is immediate, an interesting question deserving further
investigations is whether the agreement between theory and numerics remains
good even in the correlated case.

The idea behind the heterogeneous pair approximation introduced
here is fully general and its application to other dynamical models
evolving on complex networks is indeed a promising avenue for future work.
In particular, its role could be crucial for understanding cases
where heterogeneous mean-field approaches do not work~\cite{Castellano06}.

\acknowledgments
We thank A. Barrat and R. Pastor-Satorras for a critical reading
of the manuscript.

\end{document}